\documentclass[runningheads]{llncs}
\usepackage[T1]{fontenc}
\usepackage{cite}
\usepackage{graphicx}
\usepackage{multirow}
\usepackage{amsmath,amssymb,amsfonts}
\usepackage{algorithmic}
\usepackage{graphicx}
\usepackage{hyperref}
\usepackage{textcomp}
\usepackage{xcolor}
\usepackage[multiple]{footmisc}
\usepackage{bigfoot}

\DeclareNewFootnote{AAffil}[arabic]
\DeclareNewFootnote{ANote}[fnsymbol]

\begin{document}
\title{Towards reconstruction of Pulsed-wave Doppler signals from Non-invasive fetal ECG}
%
\author{Aman Verma$^*$ \inst{1}
        \and 
        Deva Satya Sriram Chintapenta$^*$ \inst{1}
        \and
        Saikat Majumder\inst{1}}
\authorrunning{A. Verma al.}
%
\institute{Department of Electronics and Communications Engineering, \\ National Institute of Technology Raipur, India \inst{1} \\
\email{aman.verma.nitrr@gmail.com,cdssriram@gmail.com,smajumder@etc.nitrr.ac.in}}
\maketitle              
\def\thefootnote{*}\footnotetext{Equal contribution}
\begin{abstract}
Fetal cardiac health monitoring with invasive methods have a limited viability because they can only be utilized during labor and are uncomfortable. 
On the other hand non-invasive fECG are adulterated with maternal ECG, and hence resulting in poor analysis. 
In contrast, Pulsed-wave Doppler (PwD) echocardiography generates high-quality signals representing fetal blood volume inflow-outflow. It also follows non-invasive signal acquisition.
The only drawback is that it requires highly expensive setup. 
To address this aspect, we put forward a challenging research question - can we reconstruct PwD signals using non-invasive fetal ECG? 
To answer this question, we perform a feasibility study with respect to input fECG wave polarity, output PwD signal configuration (EA+, EA- and group), output PwD envelopes (upper, lower, PCA compressed and group), signal length of input fECG signal and different regression models. In order to achieve good reconstruction, we also propose PwDRecNet – a deep learning framework which operates over multiple temporal contexts. 
To the best our knowledge this is the first work to consider PwD signal reconstruction from NI-fECG. The obtained numerical results suggests that with adequate configuration and model better reconstruction can be obtained.

\end{abstract}

\section{Introduction}
\label{sec:introduction}
For adults, Electrocardiogram (ECG) has been a gold-standard for cardiac activity monitoring, hence for arrhythmia prediction in fetus, fECG has been taken in account.
fECG is defined as the electrical physiological signal generated by fetal cardiovascular system \cite{mohebbian2023semi,nakatani2022fetal,rai2023fetal}. Scalp-invasive techniques have been used to measure it during labour, but this limits regular morphological fetal cardiac health monitoring. The method used to monitor fECG should both physically and economically feasible to be repeated as many numbers of times. Since invasive methods follow operation-based strategies they cannot be practiced at a prevalent scale. Non-invasive fECG (NI-fECG) has emerged as an alternative for invasive methods. It is collected via multi-electrode set-up placed over maternal abdomen. Although non-invasive, the NI-fECG signal comprises maternal ECG (MECG) and high noise \cite{wang2023correlation}. This degrades the quality of analysis. Also, it does not encapsulates information about cardiac blood inflow and outflow.

\begin{figure*}[!t]
\small
\centering
\includegraphics[height=6cm,width=1.0\linewidth]{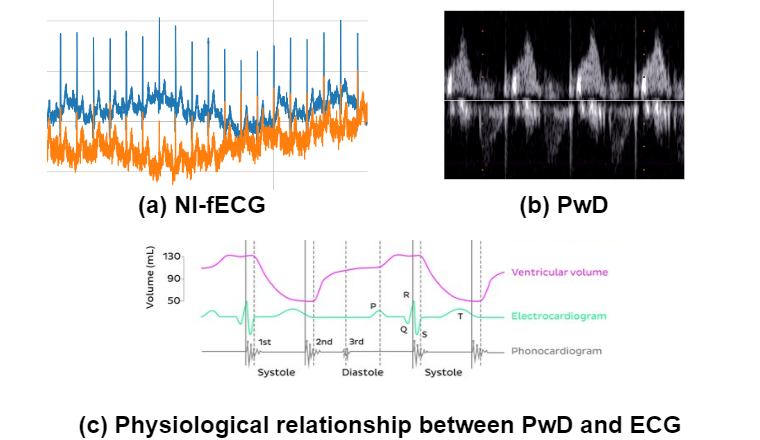}
\caption{The figure illustrates NI-fECG and PwD signals. It also shows physiological relationship between ECG and PwD for an adult ((c) part of the figure is adapted from \url{https://kenhub.com}).} 
\label{fig:PwD_Representation}
\end{figure*}

Pulsed-wave Doppler (PwD) signals on the other hand are more advanced techniques for cardiac health monitoring. It measures the blood velocity as well as the mitral inflow and aortic outflow, which conveys a richer physiological information with respect to fECG. Extraction of PwD signals involves ultrasound-based examination of fetus’s heart \cite{sulas2018automatic}. 
The spectral doppler signal does not only show blood velocity and corresponding flow volume, but also indicates the timing of cardiac events and intensity of flow (refer Fig. \ref{fig:PwD_Representation} (b)). Physiologically, PwD signal can be decomposed into atrioventricular cycles. Any such cycle can be characterized by three main waves of blood flow, when the blood flows through the mitral valve two waves are constituted: (i) \textbf{E-Wave:} This is the wave corresponding to passive filling of the ventricle due to the differential pressure between the two chambers, and (ii) \textbf{A-Wave:} This is the wave corresponding to the active filling of ventricle due to atrial contraction. These two waves represent the atrial blood inflow activity. In contrast, the aortic outflow is represented via the \textbf{V-Wave}. Although, the signal encompasses significant information regarding fetal cardiac cycles, it's acquisition requires an expensive setup. This restricts its wide-scale usage, especially in low-income countries. To this end, we find consistent patterns between ECG and ventricular volume, as illustrated in Fig. \ref{fig:PwD_Representation} (c). It is evident from the same figure that ECG uniformly changes with respect to ventricular volumetric flow of blood. This corresponds to the fact that there exists morphological correspondence between EAV-Waves and respective ECG signals. Being motivated from the same, we put forward a research question: "Can we reconstruct the PwD signal using the Non-invasive fECG signal?” By realizing this reconstruction, affordable and convenient NI-fECG signals can utilized to recover highly informative PwD signal without needing any specific setup. In order to explore the feasibility of the same, we experiment with multiple physiological settings. The results obtained are encouraging, but suggests there is requirement of lot of research thrust to enable such cross-modality reconstruction. Following are the key contributions of this work: 

\begin{itemize}
    \item We propose PwD signal envelope reconstruction using ubiquitous NI-fECG signals. To the best of our knowledge, this is the first work to explore this aspect. 
    \item In order to check upon the feasibility of the reconstruction, we conduct in-depth validation on output wave configurations, fECG polarities, timing of samples, and different models.
    \item We also propose PwDRecNet, a residually connected UNet style architecture which capture contexts at multiple-levels.
\end{itemize}

In the next section we present a literature review and then in the following section methodology is explained. Then, in subsequent section experimental Analysis is presented, while the report is concluded in the section that follows.

\section{Related works}

\subsection{PwD signal reconstruction}
As mentioned earlier, it is challenging to extract discriminative features from the NI-fECG signal due to inherent noise and multi-signal waveform. Hence, it has been a key research question – ‘How to separate fECG from MECG?’. To this end, research in \cite{mohebbian2021fetal, wang2023correlation} utilized Attention-Mechanism along with CycleGAN to achieve the state-of-the-art results. As mentioned earlier, NI-fECG does not reveal much information about blood dynamics inside the fetal heart, which in-turn limits the diagnosis of anomaly localization. Hence, usage of PwD based fetal cardiac health diagnosis seems a brighter choice. Towards, this in \cite{sulas2018automatic} authors tried to classify the cycles of PwD signals in EA+, EA- and incomplete cycle. Towards the same, authors extracted the envelopes and then used an artificial neural network for classification. The achieves performance was satisfactory but left open research gaps. In the subsequent work \cite{sulas2018fetal}, the same authors increased the dataset size and utilized similar method to validate on larger scales. Finally, in \cite{sulas2020automatic} authors gave three different approaches to classify the PwD envelopes as full cycles or incomplete cycles. They utilized fiducial methods as well as ANN based methods to realize the task. Although, there had been a recent thrust towards PwD, the field remains significantly open. There had been no methods to extract PwD signals in more economical manner, towards this end, this project presents a novel approach to alleviate the same.

\subsection{Cross-modality biosignal reconstruction}
Bio-signal reconstruction has recently caught attention, in a pilot study, authors in \cite{zhu2019ecg} tried to reconstruct ECG from more ubiquitous signal – PPG. The results obtained established a strong baseline and further motivated physiology to physiology domain adaption. In \cite{zhu2021learning}, authors further improvised the ECG reconstruction using PPG, they also gave a mathematical model and utilized different regression algorithms. 
Similar to ECG reconstruction, in \cite{zschocke2021reconstruction} pulsed-wave and respiration were reconstructed using wrist accelerometer data. 
Authors used primitive statistical models for the reconstruction and hence there remains a gap to further improvise the performance using more sophisticated regressive models as well as better training methodologies. 
Another example of bio-signal reconstruction follows from i-PPG \cite{luo2021dynamic}, wherein image-based signals are collected and with respect to contrast between RGB channels, blood flow and oxygen saturation is estimated. In \cite{comas2021turnip}, authors proposed a novel time-series UNet \cite{ronneberger2015u} for i-PPG but utilized NIR imaging techniques. 
These researches have realized cross-modality biosignal reconstruction. In particular, they perform reconstruction between biosignals which are easy to collect. Thus, there remains a need to strictly validate the reconstruction methods under practical protocols and challenging scenarios.

\section{Proposed reconstruction framework}
In order to reconstruct PwD signals from NI-fECG we follow a two-stage framework. In the first stage, we preprocess the input NI-fECG signals. Also, PwD signals are obtained as images, thus with adequate processing they are converted into time-series envelopes, to serve as ground truths. After preprocessing stage, PwDRecNet comes into play (the second stage), and reconstructs PwD signal envelopes from NI-fECG signal. We discuss each of these, in the subsequent subsections.

\subsection{Signal preprocessing}

\begin{figure*}[!t]
\small
\centering
\includegraphics[width=1.0\linewidth]{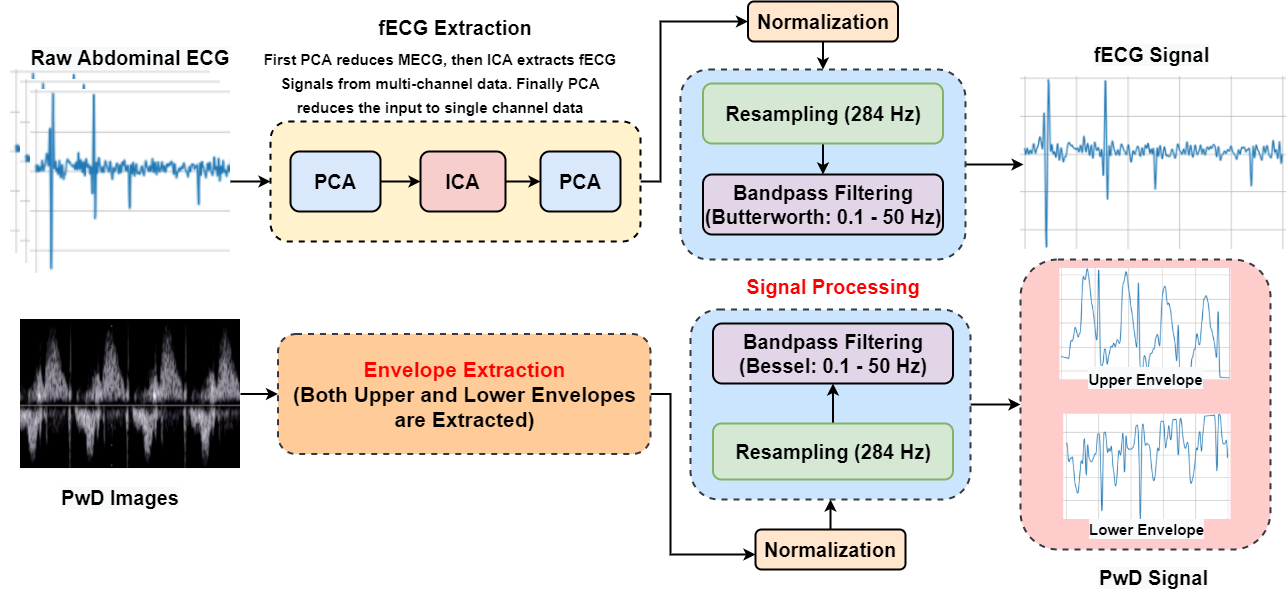}
\caption{The proposed signal preprocessing pipeline. After preprocessing, PwD signal envelopes are used as ground truths, while NI-fECG signals are used inputs for reconstruction.} 
\label{fig:PwD_SignalProcessing}
\end{figure*}

In order to enable PwD signal reconstruction several preprocessing steps at both the PwD signal as well as NI-fECG is must. The foremost reason is that these signals do not bear correspondence with respect to frequency bands and sampling rates. Further, PwD signals are made available in image formats from which envelopes must be extracted. To address these, we propose novel preprocessing pipeline (as illustrated in Fig. \ref{fig:PwD_SignalProcessing}). For the PwD signal, firstly, envelope extraction is performed. This step is performed as explained in \cite{sulas2021non}, while it involves intensity normalization, image binarization using Otsu’s Method and finally envelope extraction using max-min approach. Once, the upper and lower envelopes are extracted we perform mean normalization over them. This step also removes the DC component present in the signal. The extracted signals are then subjected to interpolation so as to idealize them at similar frequency of 284 Hz. Finally, using a Bessel bandpass Filter (0.1 – 50 Hz), denoising and removal of artefacts is performed. Filtering the signal in this range further permits to remove primary component (present at 60 Hz frequency).

NI-fECG signals are not directly available, they have to be extracted from the available AECGs which in-turn also comprises of MECG and noise. We take in input from the 3 bipolar channels (setup has been explained in [1]) and then using information contexts present at multi-channel level we extract the fECG using PCA-ICA-PCA pipeline. The first PCA operation eliminates the MECG, then ICA extracts the corresponding fECG from each signal. Characteristically, fECG is quite smaller in amplitude when compared with MECG counterpart. Then, with the final fECG signal we remove the redundant information and preserve the original three channel data by compressing it into a singular dimension. Hence, via this chain we firstly extract the fECG signal, then subsequently we perform Z-Score normalization, resampling to 284 Hz by interpolation a signal denoising (i.e., band-limiting to 0.1 – 50 Hz) via Butterworth bandpass filter. Once both the signals (PwD and fECG) are brought down to same sampling frequency, we split them in fixed time intervals. These respectively coherent signals are passed to PwDRecNet model for reconstruction training.

\subsection{PwDRecNet model}

\begin{figure*}[!t]
\small
\centering
\includegraphics[width=1.0\linewidth]{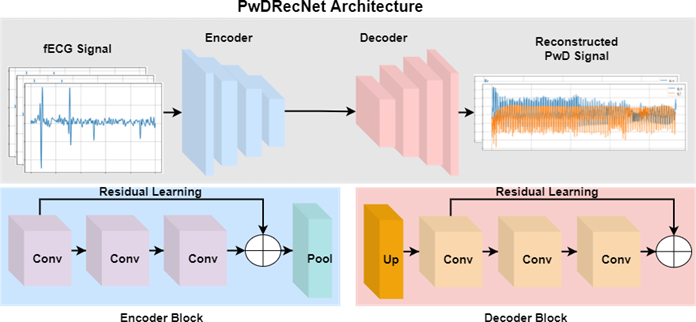}
\caption{The proposed PwDRecNet. It follows a UNet style residually connected 1D CNN architecture. It can capture temporal contexts at multiple levels. As illustrated, it's operation involves taking NI-fECG signal as input and then generating PwD signal envelopes at the output.}
\label{fig:PwD_PwDRecNet}
\end{figure*}

To enable reconstruction of envelopes of PwD signal, we propose PwDRecNet model. This model takes in processed fECG as the input, and generates corresponding PwD envelopes. The overall architecture of the PwDRecNet has been illustrated in Fig. \ref{fig:PwD_PwDRecNet}. At broader scales, PwDRecNet follows an encoder-decoder architecture, with UNet style pooling and upsampling. The encoder is constituted of three encoder-blocks which in-turn comprises of three residually connected 1D- Convolutions. These three convolutions have residual connection facilitated between them, which is instrumental in preserving temporal features as well as mitigating vanishing gradients. Finally, a pooling layer is present at the output of an encoder block. The encoded output is passed to the decoder, which follows a symmetric architecture as the encoder. There are three decoder blocks in decoder, which are composed of an upsampling layer and three residually connected 1D-Convolutions as they were in an encoder block. The output of final decoder block is made to pass through another convolution which reduces the channels dimensions to two, in order to generate respective PwD envelopes.

For training PwDRecNet, mean-squared error has been utilized while the training was performed for 50 epochs (with the best model being saved). We used RMSprop as the optimizer with 1e-3 being the learning rate. It was empirically determined that batch size of 128 gave optimal results and with respect to that batch size was set to 128.

\section{Dataset used, protocol, and performance metrics}
We perform all the experimentation over NInFEA-DB \cite{sulas2021non,goldberger2000physiobank}. It consists of data from 60 records collected from 33 pregnant women. Each record has few seconds long PwD and consists of 27 channel AECG recording. Further, the sampling rate of AECG records was 2048 Hz. All the fetus considered in the database are healthy. For conducting experimentation, we consider 80-20\% division of a record on the basis of time as well as random split. For division on the basis of time. first 80\% of all the records was used in training while the remaining in testing. Whereas, in random split setting, the complete record was randomly divided in ratio of 80\% and 20\% for training and testing respectively. We use Pearson's correlation coefficient ($r$) and mean square error ($MSE$) for evaluation. Throughout the manuscript, we will refer ($+$) and ($-$)
as the values which are respectively positive and negative, but are close to zero.

In order to evaluate the reconstruction we conduct ablation studies on: (i) input signal length and batch size, (ii) output wave configuration, (iii) output envelope selection, (iv) fECG polarity, (v) Output PwD envelope waveform, and (vi) regression models. 

\section{Experimental analysis}

\subsection{Ablation study on input signal length and batch size}
\begin{table}[!t]
\caption{Ablation Study on signal length and batch size. Results have been reported in $r$. The best result is highlighted. These results show that signals with higher temporal contexts have higher information and patterns that are intrinsically required for reconstruction.}
\centering
\label{tab:time}
\centering
\begin{tabular}{|c|ccccc|}
\hline
\multirow{2}{*}{\textbf{Input signal length (in s)}} & \multicolumn{5}{c|}{\textbf{Batch size}}                                                                                                                      \\ \cline{2-6} 
                                      & \multicolumn{1}{c|}{\textbf{32}} & \multicolumn{1}{c|}{\textbf{64}} & \multicolumn{1}{c|}{\textbf{128}}    & \multicolumn{1}{c|}{\textbf{256}} & \textbf{512} \\ \hline
\textbf{0.25}                         & \multicolumn{1}{c|}{+}           & \multicolumn{1}{c|}{+}           & \multicolumn{1}{c|}{0.0194}          & \multicolumn{1}{c|}{0.0169}       & 0.0186       \\ \hline
\textbf{0.5}                          & \multicolumn{1}{c|}{+}           & \multicolumn{1}{c|}{+}           & \multicolumn{1}{c|}{0.0121}          & \multicolumn{1}{c|}{0.0145}       & +            \\ \hline
\textbf{0.75}                         & \multicolumn{1}{c|}{-}           & \multicolumn{1}{c|}{+}           & \multicolumn{1}{c|}{0.0140}          & \multicolumn{1}{c|}{0.0139}       & 0.0156       \\ \hline
\textbf{1}                            & \multicolumn{1}{c|}{+}           & \multicolumn{1}{c|}{0.0211}      & \multicolumn{1}{c|}{-}               & \multicolumn{1}{c|}{+}            & 6            \\ \hline
\textbf{2}                            & \multicolumn{1}{c|}{0.0244}      & \multicolumn{1}{c|}{0.0236}      & \multicolumn{1}{c|}{\textbf{0.0245}} & \multicolumn{1}{c|}{0.0196}       & 0.0190       \\ \hline
\end{tabular}
\end{table}

It is essential to determine as well study the effects of signal length and batch size upon PwD envelope reconstruction, hence in this experimentation we try to achieve the same. The results of this ablation study have been tabulated in Table \ref{tab:time}. It can be observed from the same table, that the optimum results are obtained for t=2s, and specifically at batch size 128. The reason behind the same is that the signals with higher temporal contexts have higher information and patterns that are intrinsically required for reconstruction. Similarly, with higher batch size inter-sample interaction is enhanced and thus, model now examines multiple examples at once and therefore learning of amplitude-based details are exchanged. However, it is worthwhile mentioning that the achieved performance is significantly low, and the key reasons for the same are: (i) time alignment between the PwD and fECG signal is not explicitly present, (ii) fECG and PwD do not have all the signals in the same polarity, and (iii) fECG is quite challenging to recover and further in terms of amplitude it is quite low when compared with PwD.

\subsection{Ablation study on output wave configuration}
\label{sec:waveconfig}
\begin{table}[!t]
\caption{Ablation Study on wave configuration. Results have been reported in $r$. The best result is highlighted. When only EA+ configuration is reconstructed we attain gains in performance.}
\centering
\label{tab:waveconfig}
\centering
\begin{tabular}{|c|ccc|}
\hline
\multirow{2}{*}{\textbf{Wave configuration}} & \multicolumn{3}{c|}{\textbf{Time (sigal length) in seconds}}                               \\ \cline{2-4} 
                                             & \multicolumn{1}{c|}{\textbf{t=0.75}} & \multicolumn{1}{c|}{\textbf{t=1}} & \textbf{t=2}    \\ \hline
\textbf{EA+}                                 & \multicolumn{1}{c|}{0.0242}          & \multicolumn{1}{c|}{+}            & \textbf{0.0376} \\ \hline
\textbf{EA-}                                 & \multicolumn{1}{c|}{+}               & \multicolumn{1}{c|}{-}            & -               \\ \hline
\textbf{Grouped (EA+ and EA-)}               & \multicolumn{1}{c|}{0.0140}          & \multicolumn{1}{c|}{+}            & 0.0245          \\ \hline
\end{tabular}
\end{table}

Next, we analyze the behaviour of reconstruction under different configurations of PwD envelope. As mentioned in section \ref{sec:introduction}, PwD envelopes have different orientations with respect to position of fetus, this leads to reconstruction being affected as polarities of corresponding fECG does not always remains in synchronous with PwD signals’ orientation. The results of this study have been shown in Table \ref{tab:waveconfig}, we have trained the PwDRecNet model for three possible configurations (EA+, EA- and both of them together: group protocol) under three different time of signal (t=0.75,1,2). It is evident from the results that EA+ configuration of PwD is reconstructed with more ease, as for t=2s it obtains r as 0.0376. The reason behind this is that the most of the fECG signals are positively polarized, this makes the EA+ signal being at same polarity being reconstructed in better manner. In contrast, performance degradation was noticed for EA-, while the performance nearly averages out of for group protocol.

\subsection{Ablation study on output envelope selection}
\label{sec:envelope}
\begin{table}[!t]
\caption{Ablation Study on output envelope selection. The best result is highlighted. For positive envelope and EA+ enhanced performance is observed.}
\centering
\label{tab:envelope}
\centering
\begin{tabular}{|c|c|c|}
\hline
\multicolumn{1}{|l|}{\textbf{Wave configuration}} & \textbf{Envelope} & \textbf{r}      \\ \hline
\multirow{3}{*}{\textbf{EA+}}                     & Upper             & \textbf{0.0453} \\ \cline{2-3} 
                                                  & Lower             & +               \\ \cline{2-3} 
                                                  & Both              & 0.0376          \\ \hline
\multirow{3}{*}{\textbf{EA-}}                     & Upper             & 0.0227          \\ \cline{2-3} 
                                                  & Lower             & -               \\ \cline{2-3} 
                                                  & Both              & -               \\ \hline
\multirow{3}{*}{\textbf{Group}}                   & Upper             & -               \\ \cline{2-3} 
                                                  & Lower             & -               \\ \cline{2-3} 
                                                  & Both              & 0.0245          \\ \hline
\end{tabular}
\end{table}

From ablation study on wave configuration, it is clear that PwD reconstruction improves if fECG bears the same polarity. To this end, we further investigate by studying the effects over reconstruction when different configuration of PwD signals is utilized for different envelope reconstruction (i.e., Upper, Lower and both the envelopes). The obtained results have been tabulated in Table \ref{tab:envelope}. It can be observed from the table that the PwDRecNet model finds it comparatively easier to reconstruct the Upper Envelope at EA+ configuration. At this setting, it attains highest $r$ value of $0.0453$. For EA- configuration also, reconstruction of Upper envelope is relatively less challenging. The key reason behind this is that upper envelopes and fECG are aligned in similar direction. However, for an fECG which is polarized in positive direction it is quite challenging to reconstruct the lower envelope as patterns for lower envelopes are not prominently present. For the group, model it is inherently challenging to reconstruct single envelope. This arises the need to first combine the upper and lower envelope to formulate atrioventricular cycle.

\subsection{Ablation study on fECG polarity}
\begin{table}[!t]
\caption{Ablation study on fECG polarity. Results have been reported in $r$.}
\centering
\label{tab:polarity}
\centering
\begin{tabular}{|c|c|cc|}
\hline
\multirow{2}{*}{\textbf{fECG polarity}}       & \multirow{2}{*}{\textbf{Wave configuration}} & \multicolumn{2}{c|}{\textbf{Input signal length (in sec)}} \\ \cline{3-4} 
                                              &                                              & \multicolumn{1}{c|}{\textbf{t=0.75}}     & \textbf{t=2}    \\ \hline
\multirow{3}{*}{\textbf{+ve}}                 & \textbf{EA+}                                 & \multicolumn{1}{c|}{-}                   & 0.0198          \\ \cline{2-4} 
                                              & \textbf{EA-}                                 & \multicolumn{1}{c|}{-}                   & -               \\ \cline{2-4} 
                                              & \textbf{Group (EA+ and EA-)}                 & \multicolumn{1}{c|}{-}                   & -               \\ \hline
\multirow{3}{*}{\textbf{-ve}}                 & \textbf{EA+}                                 & \multicolumn{1}{c|}{-}                   & 0.0178          \\ \cline{2-4} 
                                              & \textbf{EA-}                                 & \multicolumn{1}{c|}{0.0264}              & -               \\ \cline{2-4} 
                                              & \textbf{Group (EA+ and EA-)}                 & \multicolumn{1}{c|}{-}                   & -               \\ \hline
\multirow{3}{*}{\textbf{Group (+ve and -ve)}} & \textbf{EA+}                                 & \multicolumn{1}{c|}{-}                   & -               \\ \cline{2-4} 
                                              & \textbf{EA-}                                 & \multicolumn{1}{c|}{-}                   & -               \\ \cline{2-4} 
                                              & \textbf{Group (EA+ and EA-)}                 & \multicolumn{1}{c|}{-}                   & -               \\ \hline
\end{tabular}
\end{table}

We further investigate over efficacy in reconstruction with respect to polarity of fECG, the results of the same has been tabulated in Table \ref{tab:polarity}. It can be observed from the obtained results (Table 5) that polarity do affect the reconstruction. In all of the Group models (with respect to fECG polarity and PwD Wave Configurations), reconstruction is quite challenging. The key reason behind this is that the model is not getting enough positive or negative temporal contexts to learn signal-to-signal mapping. The performance is relatively better for EA- configuration with -ve polarity of fECG and for EA+ configuration in +ve polarity of fECG. This is a representative measure of the fact that with polarity, and correct alignment of both the signals reconstruction performance can be enhanced.

\subsection{Ablation study on output PwD envelope waveform}

\begin{table}[!t]
\caption{Ablation study on output PwD envelope waveform. Results have been reported in $r$. The best attained performance has been highlighted. }
\centering
\label{tab:output}
\centering
\begin{tabular}{|c|c|cc|}
\hline
\multirow{2}{*}{\textbf{Output waveform}}                                                   & \multirow{2}{*}{\textbf{Wave configuration}} & \multicolumn{2}{c|}{\textbf{Input signal length (in sec)}} \\ \cline{3-4} 
                                                                                            &                                              & \multicolumn{1}{c|}{\textbf{t=0.75}}   & \textbf{t=2}      \\ \hline
\multirow{3}{*}{\textbf{Original}}                                                          & \textbf{EA+}                                 & \multicolumn{1}{c|}{0.0242}            & \textbf{0.0376}   \\ \cline{2-4} 
                                                                                            & \textbf{EA-}                                 & \multicolumn{1}{c|}{+}                 & -                 \\ \cline{2-4} 
                                                                                            & \textbf{Group (EA+ and EA-)}                 & \multicolumn{1}{c|}{0.0140}            & 0.0245            \\ \hline
\multirow{3}{*}{\textbf{\begin{tabular}[c]{@{}c@{}}w/ PCA\\ (Single channel)\end{tabular}}} & \textbf{EA+}                                 & \multicolumn{1}{c|}{0.0452}            & 0.0421            \\ \cline{2-4} 
                                                                                            & \textbf{EA-}                                 & \multicolumn{1}{c|}{-}                 & -                 \\ \cline{2-4} 
                                                                                            & \textbf{Group (EA+ and EA-)}                 & \multicolumn{1}{c|}{+}                 & 0.0149            \\ \hline
\end{tabular}
\end{table}

We further explore the effects of compressing the multi-channel output into a single signal, by performing PCA operation over the lower and upper envelope. The obtained results have been tabulated in Table \ref{tab:output}. When both the channels are compressed with PCA, improvement in reconstruction is observed for EA+ configuration. The reason behind this is the fact that, now the model has to generate a single output from a single input, and further the signals are more aligned. Further, performance degradation is observed for Group Model with PCA, and the possible reason is again that the signals are not properly aligned. As far as low performance in EA- configuration is concerned, it is because most of the fECGs bear positive polarity which leads to condition wherein there is very limit limited signal content with negative polarity. This forces model, being not able to reconstruct the output of negative part.

\subsection{Comparative study on different models}
\label{sec:comp}
In order to verify the PwDRecNet’s design, we have compared its performance with different regression-based counterparts (Lasso, Ridge and Linear). The results of this experiment have been tabulated in Table \ref{tab:comp}, and supremacy of PwDRecNet is clearly inferred from the same. PwDRecNet outperforms all the regression by large margins in both EA+ and Group configuration. During training the regression, loss did not optimize. This highlights the fact that simple regression models are insufficient to model the complexity of the task. Further, during this experiment we experimented on different optimizers, from this part of the study we concluded RMSprop to be the best, while Adam optimizer to be making the gradients explosive.

\begin{table}[!t]
\caption{Comparison of PwDRecNet against regression models. Results have been reported in $r$. It is clear that PwDRecNet attains significantly better performance that all regression models.}
\centering
\label{tab:comp}
\centering
\begin{tabular}{|c|c|c|}
\hline
\textbf{Model}                                                                           & \textbf{Wave configuration}  & \textbf{Input signal length (t=0.75 s)} \\ \hline
\multirow{3}{*}{\textbf{Regression}}                                                     & \textbf{EA+}                 & +                                       \\ \cline{2-3} 
                                                                                         & \textbf{EA-}                 & -                                       \\ \cline{2-3} 
                                                                                         & \textbf{Group (EA+ and EA-)} & +                                       \\ \hline
\multirow{3}{*}{\textbf{\begin{tabular}[c]{@{}c@{}}Ridge\\ Regression\end{tabular}}}     & \textbf{EA+}                 & 0.0013                                  \\ \cline{2-3} 
                                                                                         & \textbf{EA-}                 & 0.0014                                  \\ \cline{2-3} 
                                                                                         & \textbf{Group (EA+ and EA-)} & -                                       \\ \hline
\multirow{3}{*}{\textbf{\begin{tabular}[c]{@{}c@{}}Lasso\\ Regression\end{tabular}}}     & \textbf{EA+}                 & -                                       \\ \cline{2-3} 
                                                                                         & \textbf{EA-}                 & 0.0019                                  \\ \cline{2-3} 
                                                                                         & \textbf{Group (EA+ and EA-)} & 0.0017                                  \\ \hline
\multirow{3}{*}{\textbf{\begin{tabular}[c]{@{}c@{}}PwDRecNet\\ (Proposed)\end{tabular}}} & \textbf{EA+}                 & 0.0242                                  \\ \cline{2-3} 
                                                                                         & \textbf{EA-}                 & +                                       \\ \cline{2-3} 
                                                                                         & \textbf{Group (EA+ and EA-)} & 0.0140                                  \\ \hline
\end{tabular}
\end{table}

\subsection{Reconstruction results}

\begin{figure*}[!t]
\small
\centering
\includegraphics[width=1.0\linewidth]{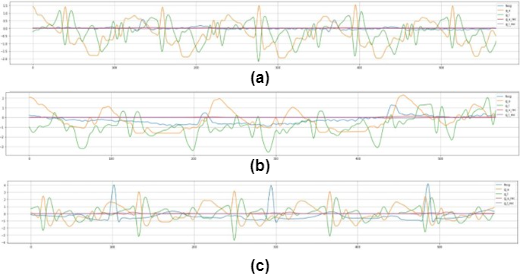}
\caption{Some reconstruction results. 'Blue' color represents the fECG singal. 'Yellow' and 'green' represents ground truth PwD signal's upper and lower envelope. While, 'maroon' and 'violet' depicts the predicted PwD signal's upper and lower envelope. It is although visually evident that reconstruction is improper, but our numerical results suggests that if proper configuration (refer Section \ref{sec:waveconfig} and \ref{sec:envelope}) and an effective model (refer Section \ref{sec:comp}) is used, then reconstruction performance can be significantly improved. This suggests that the proposed reconstruction is feasible.} 
\label{fig:PwD_results}
\end{figure*}

In this section we visually illustrate some of the reconstruction results (refer Fig. \ref{fig:PwD_results}) and then draw inferences over the reconstructed outputs. We plot fECG, ground truth PwD upper and lower envelopes and correspondingly predicted PwD envelopes. Several conclusions can be made from the same. Firstly, from all three examples (Fig. \ref{fig:PwD_results}, (a),(b) and (c)) it is evident that fECG and PwD envelopes are not aligned properly. Further, fECG is quite smaller in amplitude in comparison to the PwD signals, which are also prevalently bearing variations. Patterns and cues that must exist between fECG and PwD are also missing. Hence, during reconstruction the PwDRecNet is unable to capture contexts which leads to generation of nearly flat output (as seen in Fig. \ref{fig:PwD_results}, (a),(b) and (c)). To alleviate these issues, there is a requirement to firstly create a joint envelope using upper and lower PwD envelopes, which will bear correspondence to fECG with respect to time and periodicity. In conclusion, \textbf{\textit{our analysis suggests that if right configuration is used and a better network is developed, then reconstruction of PwD signals from NI-fECG is feasible.}}

\section{Conclusion}
This research introduced novel concept of PwD signal reconstruction using Non-invasive fECG. 
To achieve this, we proposed PwDRecNet model, a novel residually connected UNet style architecture. However, due to varying fetal heart rates, signal being misaligned in time and lack of prominent patterns, significant reconstruction has not been achieved. Extensive validation of the concept over different ablation studies over PwD wave configurations, fECG wave polarity, PwD wave envelopes and model architectures reveal data inconsistencies with respect to reconstruction. Our numerical results suggests that if proper configuration and an effective model is used, then reconstruction performance can be significantly improved. This suggests that the proposed reconstruction is feasible. To address the current shortcomings, there is a requirement to create a joint envelope using upper and lower PwD envelopes which will bear correspondence to fECG wrt and periodicity. Nevertheless, in future work we shall also be exploring better models, domain adaptation and incorporating multi-frequency contexts module to improve reconstruction performance.

\bibliographystyle{splncs04}
\bibliography{main}{}
\end{document}